\documentclass{article}

     \usepackage[final]{neurips_2019}

\usepackage[utf8]{inputenc} 
\usepackage[T1]{fontenc}    
\usepackage{hyperref}       
\usepackage{url}            
\usepackage{booktabs}       
\usepackage{amsfonts}       
\usepackage{nicefrac}       
\usepackage{microtype}      

\usepackage{graphicx}
\usepackage{amsmath,amssymb} 
\usepackage{color}
\usepackage{hyphenat}
\usepackage{subfigure}
\long\def\invis#1{}
\usepackage{multirow}
\usepackage{tipa}
\usepackage{caption}
\title{Understanding 3D CNN Behavior for Alzheimer's Disease Diagnosis from Brain PET Scan}

\author{
  Jyoti Islam \\
  Department of Computer Science\\
  Georgia State University\\
  \texttt{jislam2@student.gsu.edu} \\
  \And
Yanqing Zhang \\
Department of Computer Science \\
Georgia State University \\
   \texttt{yzhang@gsu.edu} \\
}

\begin{document}

\maketitle

\begin{abstract}
In recent days, Convolutional Neural Networks (CNN) have demonstrated impressive performance in medical image analysis. However, there is a lack of clear understanding of why and how the Convolutional Neural Network performs so well for image analysis tasks. How CNN analyzes an image and discriminates among samples of different classes are usually considered as non-transparent. As a result, it becomes difficult to apply CNN based approaches in clinical procedures and automated disease diagnosis systems. In this paper, we consider this issue and work on visualizing and understanding the decision of Convolutional Neural Network for Alzheimer's Disease (AD) Diagnosis. We develop a 3D deep convolutional neural network for AD diagnosis using brain PET scans and propose using five visualizations techniques - Sensitivity Analysis (Backpropagation),  Guided Backpropagation, Occlusion, Brain Area Occlusion,  and Layer-wise Relevance Propagation (LRP) to understand the decision of the CNN by highlighting the relevant areas in the PET data.
\end{abstract}

\section{\textbf{Introduction}}
Alzheimer's Disease (AD) is a progressive neurodegenerative disease that causes people to lose their memory, mental functions, and ability to continue daily activities. Alzheimer's Disease is primarily caused by abnormal cell death, mainly in the medial temporal lobe. Such cell death due to the accumulation of the $\beta$-amyloid peptide (A$\beta$) within the brain and the neurofibrillary tangles of hyperphosphorylated tau protein are used to identify Alzheimer's Disease [2]. Positron Emission Tomography (PET) measures the body changes at the cellular level by looking at blood flow, metabolism, neurotransmitters, and radiolabelled drugs. PET scan can identify early onset of disease before it is evident on other imaging tests [4].
\begin{figure*}[t]
\centering
\includegraphics[width=4in, height=1.5in]{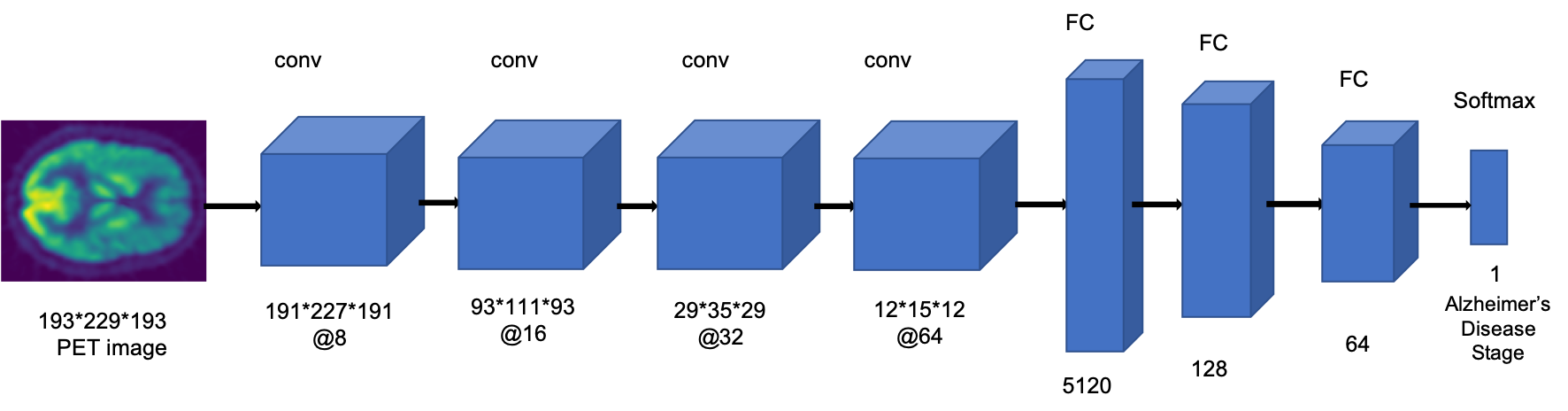}
\caption{ 3D CNN Classifier for Alzheimer's Disease diagnosis using PET data. The output of the network indicates the current stage of the patient - Normal Control (CN), Mild Cognitive Impairment (MCI) and Alzheimer's Disease (AD).}
\label{fig_cnn_3d_pet}
\end{figure*}

Deep Convolutional Neural Networks have demonstrated immense achievements for image analysis task in recent years [1], [3], [5], [7], [10], [13], [15], [17], [18] but understanding how they work yet remains a significant challenge. The decision of the Convolutional Neural Networks for image analysis is often considered non-transparent. It is essential to diagnose the disease as well as explain the reason for the diagnosis to develop a trustworthy clinical decision support system. To address this issue, understanding the behavior of CNN for classifying medical images for disease diagnosis is crucial. Computer Vision community has utilized different visualization methods to understand how Convolutional Neural Network works. Visualizing the learned features of the neurons in different layers of the network is a standard way to understand CNN behavior. A heatmap is generated and projected in the input image highlighting regions that influence the CNN for the classification decision. But there is a lack of such work for medical image analysis using convolutional neural networks. To mitigate this gap, we focus on utilizing several visualization methods to understand and explain the behavior of a convolutional neural network for Alzheimer's Disease diagnosis using brain PET data.

\section{\textbf{Methods}}
\label{proposed}
For our proposed model, we have used 1230 PET scans (169 AD, 661 MCI, 400 CN) of 988 patients. We collected the data from the  Alzheimer's  Disease Neuroimaging  Initiative  (ADNI)  database  (adni.loni.usc.edu). To ensure the relative alignment among the subjects, we non-linearly registered the PET scans to the 1mm resolution 2009c version of the ICBM152 reference brain using Advanced Normalization Tools (ANTs3).  Since the raw scans are not skull-stripped and have unnecessary information, skull stripping and normalization was performed on them. Skull stripping includes removal of non-cerebral tissues like skull, scalp, and dura from brain images and helps reduce computational complexity and time. After this step, all the PET scans resulted in volumes of 193 * 229 * 193.
We developed a 3D Convolutional Neural Network inspired by the architecture [6] for Alzheimer's Disease diagnosis. The CNN architecture is depicted in Figure 1.
To understand the decision of the CNN of AD diagnosis, we have applied five visualizations techniques - Sensitivity Analysis ( Backpropagation) [8], Guided Backpropagation [9], Occlusion [11], Brain Area Occlusion [12], and Layer-wise Relevance Propagation (LRP)  [14].
\section{\textbf{Experiments and Results}}
\label{exp}
For our work, we used 80\% data from the dataset as the training set,  and 20\% as test dataset. From the training dataset, a random selection of 10\% images is used as validation dataset. The training-test and training-validation split were done at the subject/patient level. The experiments were performed using the PyTorch framework. The parameters used for the training process are learning rate: 0.0001, weight decay: 0.1  after every seven epochs, and batch size: 16. The network achieves a comparable classification accuracy of 88.76\% for CN/AD classification. Please note that the focus of the current study is CNN visualization and not classification performance. In our experiment, we also developed a 2D-CNN model using axial, coronal, and sagittal slices from PET data that achieved 71.45\% classification accuracy for CN/AD classification. The vast difference in the classification result suggests that 3D-CNN networks have better capability to learn features from three-dimensional PET image data. Though to validate such findings, further experiments are necessary.
\begin{figure*}[t]
\begin{center}
\leavevmode
\begin{tabular}{ccc}
\includegraphics[width=.3\linewidth,height=1.25in]{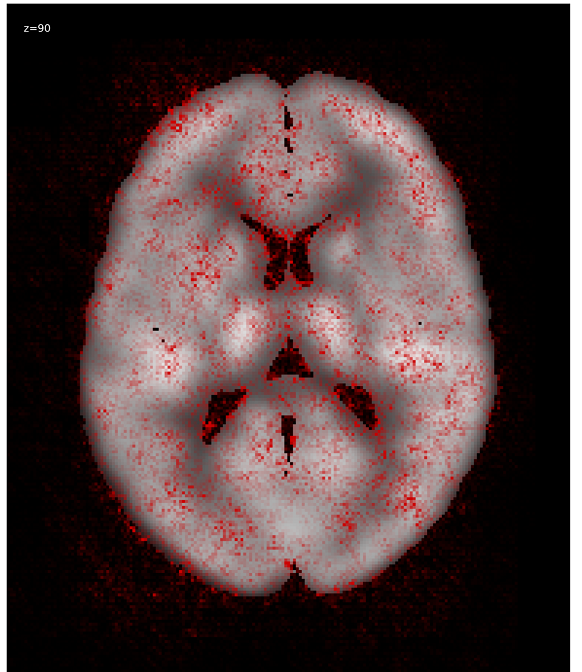} &
\includegraphics[width=.3\linewidth,height=1.25in]{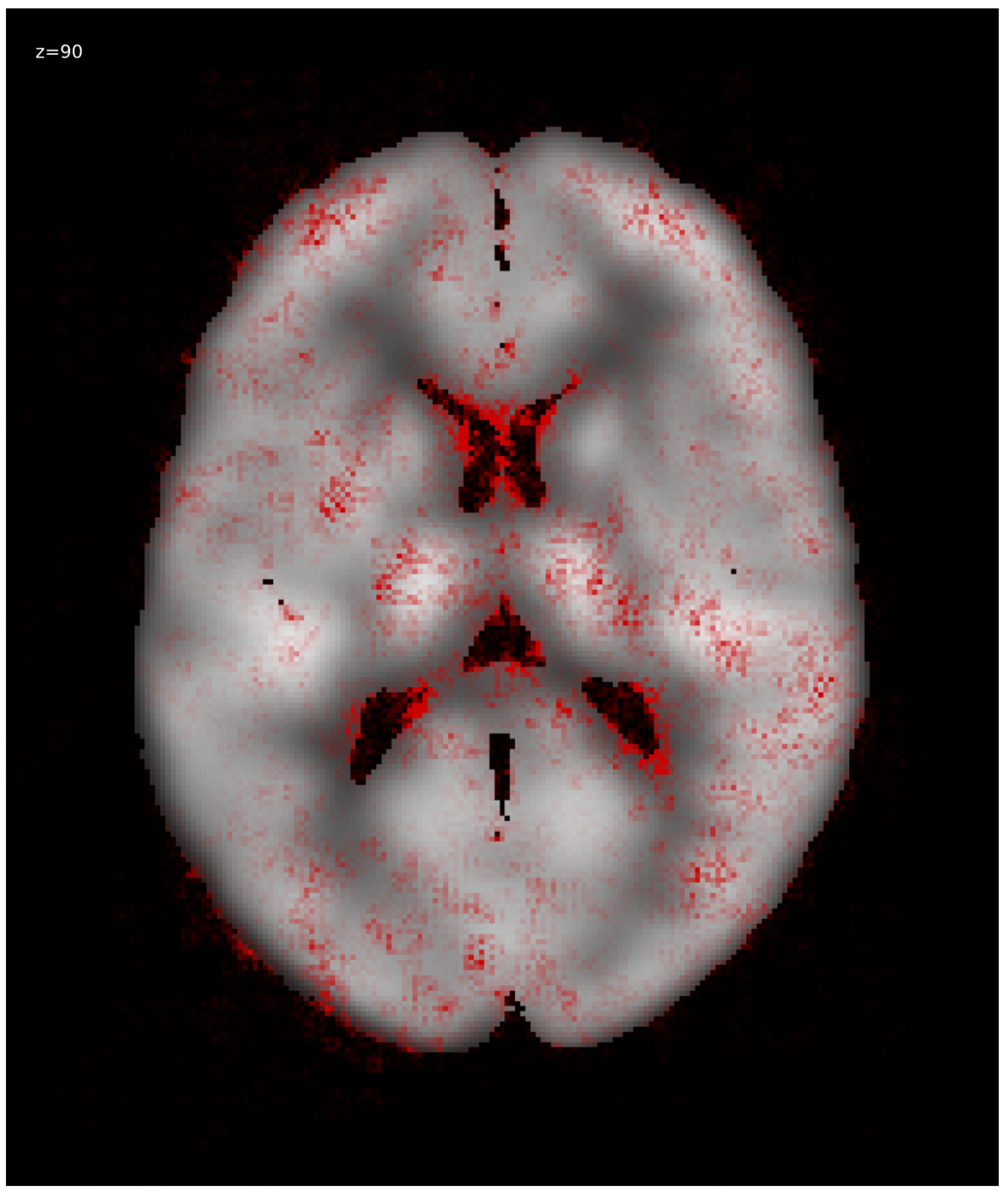} &
\includegraphics[width=.3\linewidth,height=1.25in]{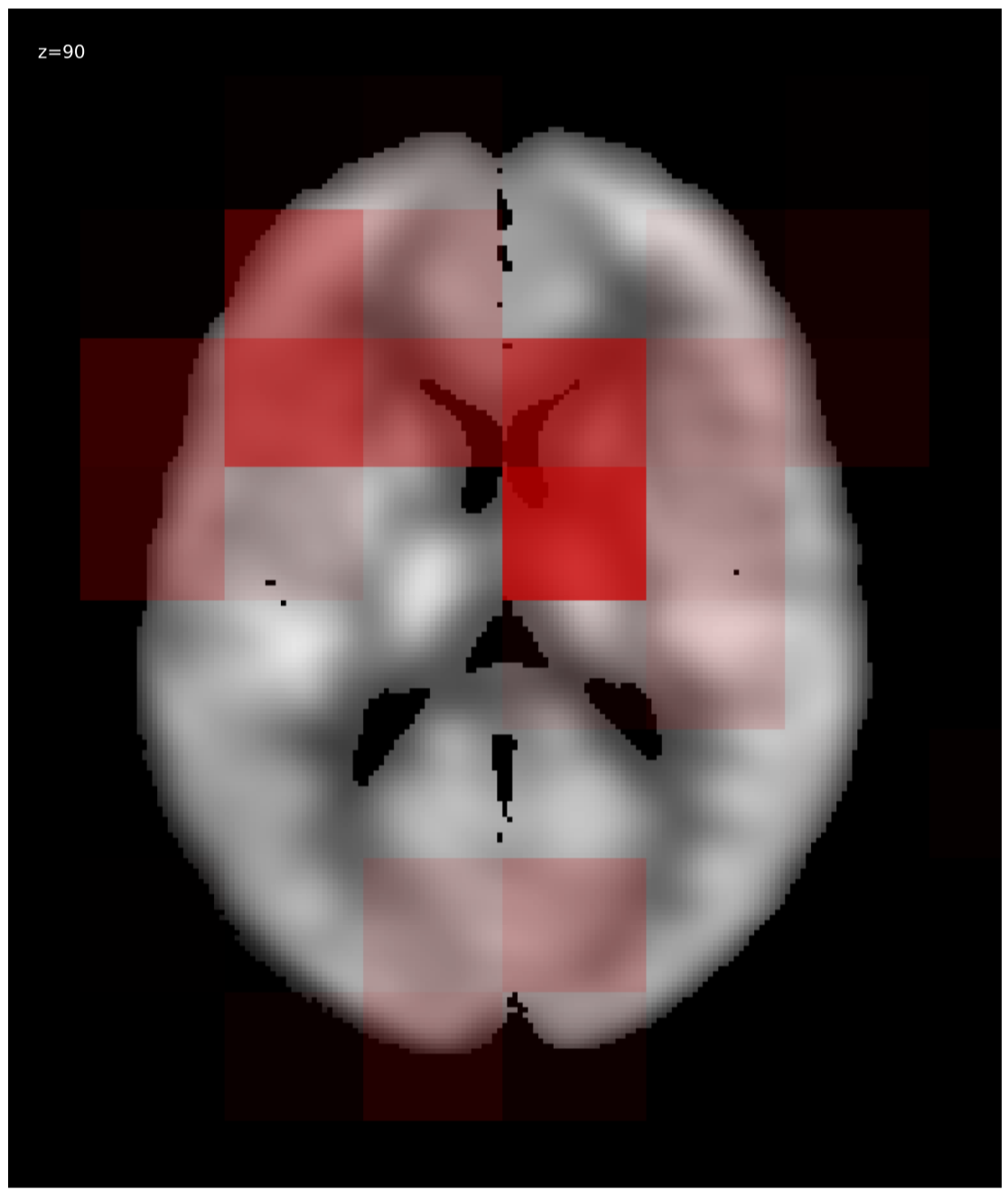} \\
(a) & (b) & (c)\\
\end{tabular}
\begin{tabular}{cc}
\includegraphics[width=.3\linewidth,height=1.25in]{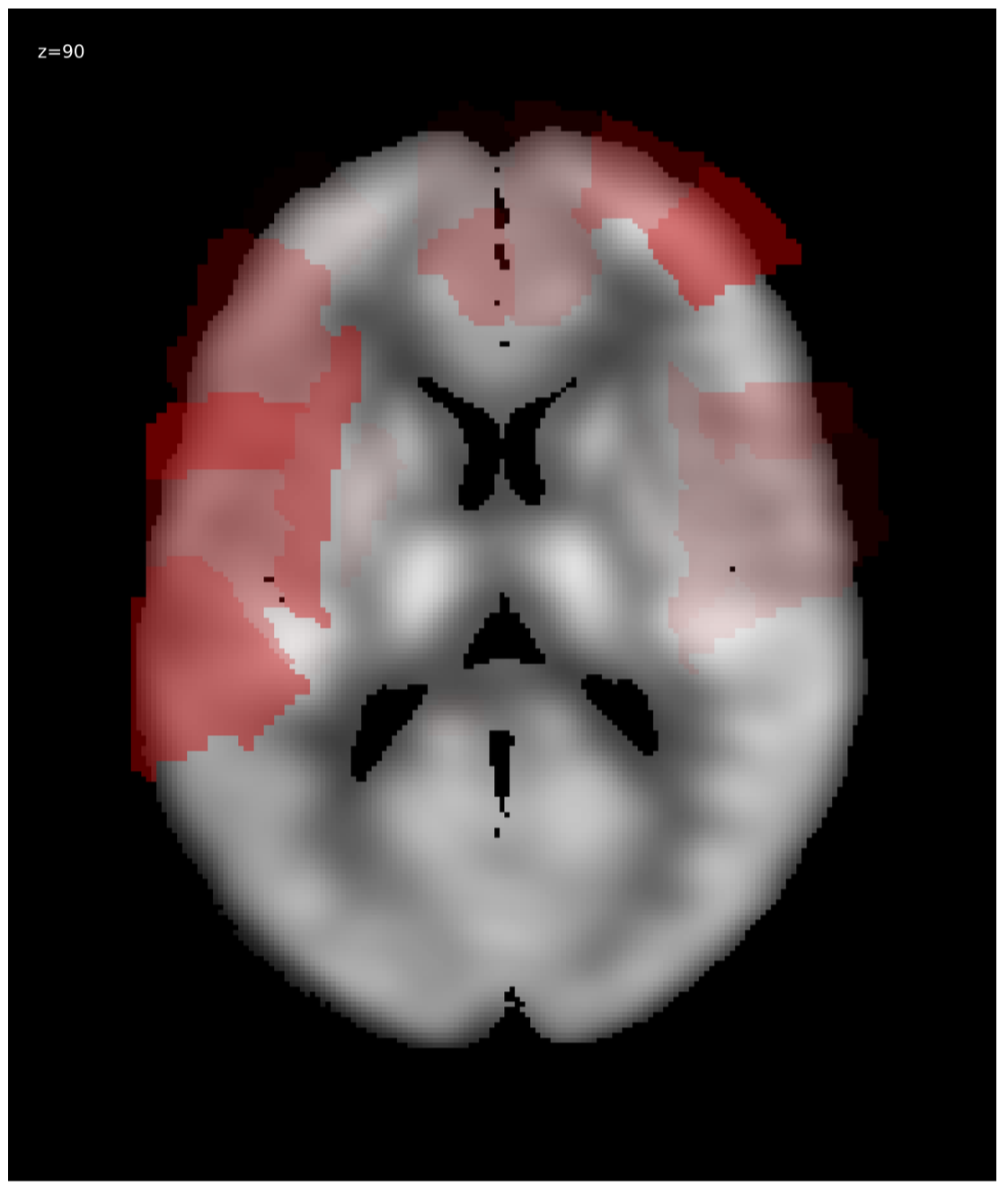} &
\includegraphics[width=.3\linewidth,height=1.25in]{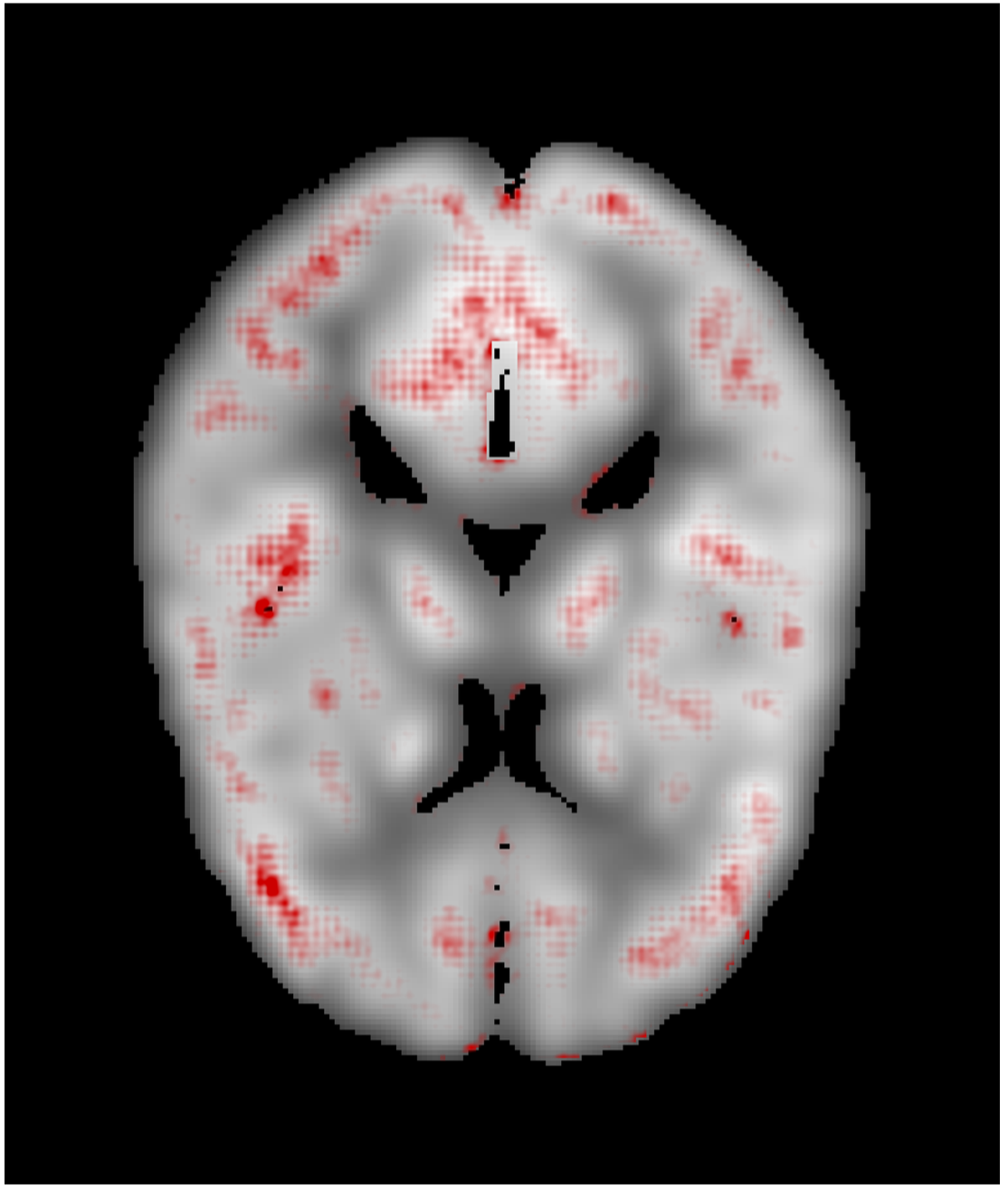} \\
(d) & (e)\\
\end{tabular}
\end{center}
\caption{Visualization comparison of the Relevance Heatmaps. (a) Sensitivity Analysis ( Backpropagation) [8]. (b) Guided Backpropagation [9]. (c) Occlusion [11]. (d) Brain Area Occlusion [12]. (e) Layer-wise Relevance Propagation (LRP)  [14].}
\label{fig_pet_comp}
\end{figure*}
\subsection{\textbf{Visualization}}

We generated relevance heatmaps for all visualizations methods, averaged over AD  PET images in the test dataset. Figure 2 presents the visual comparison of these five methods. The red areas/dots indicate that regions were important for the decision making of the 3D-CNN model. From the result, we can see that all the visualization focuses mostly on similar brain regions. There are some differences, such as the heatmaps generated for the gradient-based methods are distributed. The heatmaps highlight the areas that the CNN network is most susceptible. For the LRP method, the heatmap shows the average relevance of each voxel for contributing to the AD diagnosis score. The heatmaps generated by the occlusion based methods are more focused on the specific regions and cannot administer with large areas of distributed relevance. The reason behind the issue is the occlusion path was not able to cover those areas (for example, the cortex) completely. Brain area occlusion presents very high relevance for the temporal lobe. Since in this method, only one area is covered at a time, that can cause such high importance for one region and minimal relevance for other regions.
\subsection{\textbf{Impact of Brain Region}}
To evaluate the visualization result, we consider the literature available from the medical domain [16], [19], as there is no ground truth for validating the generated heatmaps. Table. 1  presents the top five most relevant regions for each visualization method.  The relevance in each area following the AAL atlas was summed together to identify the most relevant areas for AD PET scans. From the top five relevant brain areas for each visualization method, we can see that the 3D-CNN focuses on the temporal lobe area, including the hippocampus for CN/AD classification. So, our findings are similar to those in the medical literature [16], [19].

\begin{table*}
\centering
\caption{Brain Area Relevance for AD Patient for Each Visualization Method}
\label{tab:brainrel}
\resizebox{\textwidth}{!}{%
\begin{tabular}{|c|l|l|l|l|}
\hline
    \multicolumn{1}{|c|}{\begin{tabular}[c]{@{}c@{}}Sensitivity\\ Analysis\\ (Backpropagation)\end{tabular}} & \multicolumn{1}{c|}{\begin{tabular}[c]{@{}c@{}}Guided \\ Backpropagation\end{tabular}} &
    \multicolumn{1}{c|}{\begin{tabular}[c]{@{}c@{}}Occlusion \\ Sensitivity\end{tabular}} & \multicolumn{1}{c|}{\begin{tabular}[c]{@{}c@{}}Brain Area\\ Occlusion\end{tabular}} & \multicolumn{1}{c|}{\begin{tabular}[c]{@{}c@{}}Layer-wise Relevance \\ Propagation (LRP)\end{tabular}} \\ \hline
 \begin{tabular}[c]{@{}l@{}}
     Frontal\_Mid (5.59\%) \\
    Precuneus (4.87\%) \\
    Postcentral (4.82\%) \\
    Temporal\_Mid (4.57\%) \\
    Precentral (4.32\%)
      \end{tabular}
    & \begin{tabular}[c]{@{}l@{}}
    Frontal\_Mid (5.92\%) \\
    Postcentral (4.72\%) \\
    Precuneus (4.55\%) \\
    Precentral (4.34\%) \\
    Temporal\_Inf (3.90\%)
      \end{tabular}
    & \begin{tabular}[c]{@{}l@{}}
    Postcentral (8.77\%) \\
    Precuneus (8.14\%) \\
    Precentral (5.05\%) \\
    Parietal\_Sup (5.01\%) \\
    Temporal\_Mid (4.82\%)
    \end{tabular}
        & \begin{tabular}[c]{@{}l@{}}
     Precuneus (14.11\%) \\
    Temporal\_Mid (13.27\%) \\
    Temporal\_Inf (9.51\%) \\
    Frontal\_Mid (5.51\%) \\
    Angular (4.79\%)
      \end{tabular}
    &  \begin{tabular}[c]{@{}l@{}}
   Cerebral white matter (32.60\%) \\
    Cerebellum (9.78\%) \\
    Frontal\_Mid (3.26\%) \\
    Lateral Ventricles (3.02\%) \\
    Brain Stem (2.93\%)
    \end{tabular}                                                             \\ \hline
\end{tabular}
  }
\end{table*}

\section{\textbf{Conclusion}}
\label{concl}
We developed a 3D Convolutional Neural Network for Alzheimer's Disease Diagnosis and applied five visualization methods to understand the network's behavior to classify sample brain PET data. To the best of our knowledge, our work is the first to comprehend CNN behavior for  Alzheimer's Disease diagnosis trained using PET data. It can be extended to understanding CNN behavior for other disease diagnosis systems. The 3D-CNN focuses on the temporal lobe area, including the hippocampus for AD diagnosis. Our findings are similar to those in medical literature, and these biomarkers were identified directly from the network's decision without having any ground truth. Such findings are crucial for building trust in clinical practitioners for adapting automated disease diagnosis systems.
\section{\textbf{Reference}}

[1] J. Islam, Y. Zhang, ``Visual sentiment analysis for social images using transfer learning approach." 2016 IEEE International Conferences on Big Data and Cloud Computing (BDCloud), Social Computing and Networking (SocialCom), Sustainable Computing and Communications (SustainCom)(BDCloud-SocialCom-SustainCom), IEEE, 2016.

[2] C. Marcus, E. Mena, and R. M. Subramaniam, “Brain pet in the
diagnosis of alzheimers disease,” Clinical nuclear medicine, vol. 39,
no. 10, p. e413, 2014.

[3] Islam, J., Zhang, Y., ``Early Diagnosis of Alzheimer's Disease: A Neuroimaging Study with Deep Learning Architectures." Proceedings of the IEEE Conference on Computer Vision and Pattern Recognition Workshops. 2018.

[4] A. Berger, “How does it work?: Positron emission tomography,” BMJ:
British Medical Journal, vol. 326, no. 7404, p. 1449, 2003.

[5] Islam, J., Zhang, Y., ``Towards Robust Lung Segmentation in Chest Radiographs with Deep Learning''. In Machine Learning for Health Workshop at Conference on Neural Information Processing Systems (NeurIPS 2018), Dec. 2018.

[6] A. Khvostikov, K. Aderghal, J. Benois-Pineau, A. Krylov, and G. Catheline, “3d cnn-based classification using smri and md-dti images for
alzheimer disease studies,” arXiv preprint arXiv:1801.05968, 2018.

[7] Islam, J., Zhang, Y., ``An Ensemble of Deep Convolutional Neural Networks for Alzheimer's Disease Detection and Classification." arXiv preprint arXiv:1712.01675 (2017).

[8] K. Simonyan, A. Vedaldi, and A. Zisserman, “Deep inside convolutional
networks: Visualising image classification models and saliency maps,”
arXiv preprint arXiv:1312.6034, 2013.

[9] J. T. Springenberg, A. Dosovitskiy, T. Brox, and M. Riedmiller,
“Striving for simplicity: The all convolutional net,” arXiv preprint
arXiv:1412.6806, 2014.

[10] Islam, J., Zhang, Y., ``Deep Convolutional Neural Networks for Automated Diagnosis of Alzheimer’s Disease and Mild Cognitive Impairment Using 3D Brain MRI." International Conference on Brain Informatics. Springer, Cham, 2018.

[11] M. D. Zeiler and R. Fergus, “Visualizing and understanding convolutional networks,” in European conference on computer vision. Springer,
2014, pp. 818–833.

[12] J. Rieke, F. Eitel, M. Weygandt, J.-D. Haynes, and K. Ritter, “Visualizing
convolutional networks for mri-based diagnosis of alzheimers disease,”
in Understanding and Interpreting Machine Learning in Medical Image
Computing Applications. Springer, 2018, pp. 24–31.

[13] Islam, J., Zhang, Y., ``Brain MRI analysis for Alzheimer's disease diagnosis using an ensemble system of deep convolutional neural networks." Brain informatics 5, no. 2 (2018): 2.

[14] S. Bach, A. Binder, G. Montavon, F. Klauschen, K.-R. Muller, and
W. Samek, “On pixel-wise explanations for non-linear classifier decisions by layer-wise relevance propagation,” PloS one, vol. 10, no. 7, p.
e0130140, 2015.

[15] Islam, J., Zhang, Y., ``A novel deep learning based multi-class classification method for Alzheimer's disease detection using brain MRI data." International Conference on Brain Informatics. Springer, Cham, 2017.

[16] M. J. West, P. D. Coleman, D. G. Flood, and J. C. Troncoso, ``Differences
in the pattern of hippocampal neuronal loss in normal ageing and
alzheimer’s disease," The Lancet, vol. 344, no. 8925, pp. 769–772, 1994.

[17] Modasshir, M., Li, A. Q, and Rekleitis, I., "MDNet: Multi-Patch Dense Network for Coral Classification." OCEANS 2018 MTS/IEEE Charleston. IEEE, 2018.

[18] Modasshir, M., Rahman, S., Youngquist, O. and Rekleitis, I., ``Coral Identification and Counting with an Autonomous Underwater Vehicle." 2018 IEEE International Conference on Robotics and Biomimetics (ROBIO). IEEE, 2018.

[19] K. Jin, A. L. Peel, X. O. Mao, L. Xie, B. A. Cottrell, D. C. Henshall, and
D. A. Greenberg, ``Increased hippocampal neurogenesis in alzheimer’s
disease," Proceedings of the National Academy of Sciences, vol. 101,
no. 1, pp. 343--347, 2004.
\end{document}